\pdfoutput=1
\documentclass[12pt]{article}
\usepackage[utf8]{inputenc}
\usepackage[british]{babel}
\usepackage{cmap}
\usepackage{lmodern}

\usepackage{amssymb, amsmath, amsthm}
\usepackage[a4paper,top=25mm,bottom=25mm,left=25mm,right=25mm]{geometry}
\usepackage{ragged2e}

\usepackage{authblk} 
\usepackage{pifont}
\usepackage{graphicx}
\usepackage[dvipsnames,svgnames,table]{xcolor}
\usepackage[figuresright]{rotating}
\usepackage{xtab} 
\usepackage{longtable} 
\usepackage{multirow}
\usepackage{footnote}
\usepackage[stable]{footmisc}
\usepackage{chngpage} 
\usepackage{pdflscape} 
\usepackage[nottoc,notlot,notlof]{tocbibind} 

\usepackage{pgfplots}
\pgfplotsset{compat=1.18}
\pgfplotsset{every tick label/.append style={font=\footnotesize}}
\usepackage{setspace}

\makesavenoteenv{tabular}
\usepackage{tabularx}
\usepackage{booktabs}
\usepackage{threeparttable} 
\usepackage[referable]{threeparttablex} 
\newcolumntype{R}{>{\raggedleft\arraybackslash}X}
\newcolumntype{L}{>{\raggedright\arraybackslash}X}
\newcolumntype{C}{>{\centering\arraybackslash}X}
\newcolumntype{A}{>{\columncolor{gray!25}}C}
\newcolumntype{a}{>{\columncolor{gray!25}}c}

\newlength{\tablen}

\usepackage{dcolumn} 
\newcolumntype{.}{D{.}{.}{-1}}

\usepackage{tikz}
\usetikzlibrary{arrows, calc, matrix, patterns, positioning, shapes, trees}
\usepackage[semicolon]{natbib}
\usepackage[hyphens]{url}
\usepackage{hyperref} 
\hypersetup{
  colorlinks   = true,    
  urlcolor     = blue,    
  linkcolor    = blue,    
  citecolor    = ForestGreen      
}
\usepackage{microtype}
\usepackage[singlelinecheck=false,justification=centering]{caption} 

\usepackage[labelformat=simple]{subcaption}

\DeclareCaptionLabelFormat{parenthesis}{(#2)}
\makeatletter
\renewcommand\p@subfigure{\arabic{figure}.}
\makeatother

\DeclareCaptionLabelFormat{parenthesis}{(#2)}
\makeatletter
\renewcommand\p@subtable{\arabic{table}.}
\makeatother

\usepackage{enumitem}
\setlist[itemize]{leftmargin=2.5\parindent}
\setlist[enumerate]{leftmargin=2.5\parindent}

\def\addlegendimage{\csname pgfplots@addlegendimage\endcsname}

\theoremstyle{plain}

\theoremstyle{definition}

\theoremstyle{remark}

\def\keywords{\vspace{.5em} 
{\noindent \textit{Keywords}: }}

\def\JEL{\vspace{.5em} 
{\noindent \textbf{\emph{JEL} classification number}: }}

\def\AMS{\vspace{.5em} 
{\noindent \textbf{\emph{MSC} class}: }}

\title{Mitigating the risk of tanking \\ in multi-stage tournaments}
\author{\href{https://sites.google.com/view/laszlocsato}{L\'aszl\'o Csat\'o}\thanks{~E-mail: \emph{laszlo.csato@sztaki.hun-ren.hu}} }
\affil{Institute for Computer Science and Control (SZTAKI) \\
Hungarian Research Network (HUN-REN) \\
Laboratory on Engineering and Management Intelligence \\
Research Group of Operations Research and Decision Systems}
\affil{Corvinus University of Budapest (BCE) \\
Institute of Operations and Decision Sciences \\
Department of Operations Research and Actuarial Sciences}
\affil{Budapest, Hungary}
\date{\today}

\def\Dedication{
{\noindent
$\mathfrak{Eben}$ $\mathfrak{diese}$ $\mathfrak{Mannigfaltigkeit}$ $\mathfrak{der}$ $\mathfrak{Gegenst\ddot{a}nde}$ $\mathfrak{entsteht}$ $\mathfrak{bei}$ $\mathfrak{der}$ $\mathfrak{Pr\ddot{u}fung}$ $\mathfrak{der}$ $\mathfrak{Mittel}$, $\mathfrak{je}$ $\mathfrak{h\ddot{o}her}$ $\mathfrak{man}$ $\mathfrak{mit}$ $\mathfrak{dem}$ $\mathfrak{Standpunkt}$ $\mathfrak{hinaufr\ddot{u}ckt}$; $\mathfrak{denn}$ $\mathfrak{je}$ $\mathfrak{h\ddot{o}her}$ $\mathfrak{die}$ $\mathfrak{Zwecke}$ $\mathfrak{liegen}$, $\mathfrak{um}$ $\mathfrak{so}$ $\mathfrak{gr\ddot{o}\ss er}$ $\mathfrak{ist}$ $\mathfrak{die}$ $\mathfrak{Zahl}$ $\mathfrak{der}$ $\mathfrak{Mittel}$, $\mathfrak{welche}$ $\mathfrak{zu}$ $\mathfrak{ihrer}$ $\mathfrak{Erreichung}$ $\mathfrak{angewendet}$ $\mathfrak{werden}$.\footnote{~``\emph{The same multiplicity of circumstances is presented also in the examination of the means the higher our point of view, for the higher the object is situated, the greater must be the number of means employed to reach it.}'' (Source: Carl von Clausewitz: \emph{On War}, Book 2, Chapter 5 [Criticism]. Translated by Colonel James John Graham, London, N. Tr\"ubner, 1873. \url{http://clausewitz.com/readings/OnWar1873/TOC.htm})}
}
\vspace{0.25cm}

\flushright
\noindent (Carl von Clausewitz: \emph{Vom Kriege})

\vspace{1cm} 
\justify }

\begin{document}

\newgeometry{top=20mm,bottom=20mm,left=25mm,right=25mm}

\maketitle
\thispagestyle{empty}
\Dedication

\begin{abstract}
\noindent
Multi-stage tournaments consisting of a round-robin group stage followed by a knockout phase are ubiquitous in sports. However, this format is incentive incompatible if at least two teams from a group advance to the knockout stage where the brackets are predetermined. A model is developed to quantify the risk of tanking in these contests. The suggested approach is applied to the 2022 FIFA World Cup to uncover how its design could have been improved by changing group labelling (a reform that has received no attention before) and the schedule of group matches. Scheduling is found to be a surprisingly weak intervention compared to previous results on the risk of collusion in a group. The probability of tanking, which is disturbingly high around 25\%, cannot be reduced by more than 3 percentage points via these policies. Tournament organisers need to consider more fundamental changes against tanking.


\AMS{90-10, 90B90, 91B14}

\JEL{C44, C63, Z20}
\end{abstract}

\clearpage
\restoregeometry

\section{Introduction} \label{Sec1}

Several sports tournaments are organised in a hybrid format: a preliminary stage played in round-robin groups is followed by a knockout phase. Usually, the top $\ell$ teams in each group advance to the knockout stage, and higher-ranked teams face lower-ranked teams in the first knockout round. Group rank is supposed to reflect team abilities, that is, the group winners are on average stronger than the runners-up. Consequently, every team is interested in winning if its opponent in the first knockout round is chosen randomly such as in the UEFA Champions League, the most prestigious European club football competition.

However, in the case of predetermined knockout brackets, the necessary and sufficient condition for incentive compatibility is to allow only the top-ranked team to qualify from each group \citep{Vong2017}. Otherwise, the runner-up in group $i$ can be stronger than the winner due to bad luck, and the teams playing in other groups want to face the winner of group $i$ rather than the runner-up. Since $\ell \geq 2$ in almost all real-world tournaments, they are vulnerable to \emph{tanking}, the act of deliberately dropping points or losing a game to gain an advantage.

There are a number of instances when this problem led to serious issues of unfairness. Perhaps the most famous example is the women's doubles badminton tournament of the 2012 Olympic Games \citep[Section~3.3.1]{KendallLenten2017}. Here, four teams were ejected because of ``\emph{not using one's best efforts to win a match}'' and ``\emph{conducting oneself in a manner that is clearly abusive or detrimental to the sport}''. They were accused of trying to lose in order to get (perceived) easier draws in the knockout stage of the competition.

Analogously, the Indonesian defender \emph{Mursyid Effendi} deliberately scored an own goal during a soccer game between Thailand and Indonesia at the 1998 AFF championship \citep[Section~3.9.2]{KendallLenten2017}. Both teams already secured a place in the semifinals but the group winner would have to play against the host Vietnam in another city, while the runner-up would have to play against Singapore without travelling, which was seen as the better option. The F\'ed\'eration Internationale de Football Association (FIFA) fined both teams for ``\emph{violating the spirit of the game}'' and banned the Indonesian player for life from international football.

Naturally, it is almost impossible to prove tanking but, analogously, there is no way for the teams to verify that they played honestly. Hence, the mere existence of a possibility for tanking can be detrimental to the integrity of sports, and may lead to accusations from the other teams and spectators if the result of the game benefits a team that had an incentive for tanking.
Thus, it is important to reduce the threat of tanking to the extent possible if it can be achieved by a minimal change in the competition design. 
The current paper will examine this issue in a simulation framework, which is a standard approach in the literature, although it does not take present-day performances and psychological aspects into account.

The novelty of our study resides in the consideration of match fixing across groups, rather than from a within group perspective as previous papers do \citep{ChaterArrondelGayantLaslier2021, Csato2022a, Guyon2020a, Stronka2020, Stronka2024}. This makes it possible to analyse the group labelling rule, a potential decision directly after the group draw that has received no attention before.

The main contributions can be summarised as follows:
\begin{itemize}
\item
We develop a framework to quantify the risk of tanking in hybrid contests with a predetermined knockout bracket, a format used in several international sports tournaments such as the FIFA World Cup, the UEFA European Championship, the FIBA Basketball World Cup, or the IHF World Handball Championship.
\item
The role of the sequence of group matches is small with respect to tanking opportunities, easily dominated by the impact of the order of the last round in the groups that are paired in the knockout bracket. This finding is in stark contrast with the case of collusion within a group, where scheduling has a considerable effect on the quality of games \citep{ChaterArrondelGayantLaslier2021, Guyon2020a, Stronka2020}.
\item
According to our simulations based on the 2022 FIFA World Cup, the proposed changes in group labelling and match scheduling, are not able to decrease the disturbingly high probability of tanking, which is about 25\%, by more than 3 percentage points.
\end{itemize}
The proposed approach can be used by tournament designers to evaluate alternative schedules at the time of the group draw, and choose an option that minimises the probability of matches with misaligned incentives. However, more radical interventions---such as dynamic scheduling \citep{GuajardoKrumer2023, KrumerMegidishSela2023, Stronka2024}, opponent choice \citep{Guyon2020a, HallLiu2024}, or randomised tie-breaking \citep{Stronka2024}---would be necessary against tanking.

The remainder of the paper is organised as follows.
Section~\ref{Sec2} provides a concise literature overview. The methodology is presented in Section~\ref{Sec3}, and the results for the 2022 FIFA World Cup are provided in Section~\ref{Sec4}. Finally, Section~\ref{Sec5} concludes.

\section{Related literature} \label{Sec2}

Tanking is a standard topic of academic research since it is felt as being against the spirit of the game. \citet{KendallLenten2017} discuss a number of historical examples from various sports. In particular, the traditional player draft used to ensure competitive balance in major North American and Australian sports creates perverse incentives to lose after a team is eliminated from the playoff \citep{Fornwagner2019, PriceSoebbingBerriHumphreys2010, TaylorTrogdon2002}. Many solutions have been proposed to mitigate this problem \citep{BanchioMunro2021, Gold2010, KazachkovVardi2020, Lenten2016, LentenSmithBoys2018}.

Inspired by the women's doubles badminton tournament of the 2012 Olympics, \citet{Pauly2014} suggests a mathematical model of strategic manipulation in complex sports competitions. An impossibility theorem is proved to demonstrate that strategy-proofness cannot hold under some reasonable constraints. As already mentioned in the Introduction, \citet{Vong2017} gives the necessary and sufficient condition for incentive compatibility in multi-stage tournaments: only the group winner can qualify for the next stage.

Tanking can emerge not only because an advantage is gained in expected terms but also unintentionally, due to misaligned tournament rules. \citet{DagaevSonin2018} reveal how incentive compatibility can be guaranteed in qualification systems composed of one round-robin and several knockout tournaments. Neglecting this result has led to problems in some competitions organised by the Union of European Football Associations \citep{Csato2021a, HaugenKrumer2021}. Analogously, sports tournaments with multiple group stages, where the results of matches already played in the previous round against teams in the same group are carried over, are vulnerable to tanking \citep{Csato2022c}.

However, these studies consider incentive compatibility (IC) as a binary concept and do not attempt to measure the degree of its violation. This is a severe limitation because ``\emph{devising a method to quantify IC would allow for the study of trade-offs between IC and other potentially desirable factors}'' \citep[p.~567]{Vong2017}.
\citet{Csato2022a} presents probably the first method to quantify IC through the example of the European Qualifiers for the 2022 FIFA World Cup.

Our paper has strong connections to contest design, which has many aspects that can be addressed with the tools of operations research \citep{DevriesereCsatoGoossens2024}. These include the choice of the tournament format \citep{GoossensBelienSpieksma2012, LasekGagolewski2018, ScarfYusofBilbao2009, SziklaiBiroCsato2022}, the classification (seeding) rules of teams into divisions/groups \citep{CeaDuranGuajardoSureSiebertZamorano2020, ScarfYusof2011}, the draw mechanism used to allocate the teams into groups \citep{BoczonWilson2023, Guyon2015a, LalienaLopez2019, RobertsRosenthal2024}, the tie-breaking rules \citep{Berker2014, Csato2023a}, the prize allocation scheme \citep{DietzenbacherKondratev2023}, the scheduling of the matches \citep{RasmussenTrick2008, RibeiroUrrutiadeWerra2023a}, or the scoring method applied \citep{KondratevIanovskiNesterov2023, VaziriDabadghaoYihMorin2018}.

In particular, a recent line of literature deals with the issue of how the order of matches influences the probability of collusion and match-fixing.
\cite{Guyon2020a} calculates the risk of collusion as a function of the schedule in groups of three teams, which was planned to be used in the 2026 FIFA World Cup.
\citet{Stronka2020} introduces the so-called unanimity pair matching method, which matches the winners of two adjacent groups if they unanimously express their preference for playing against each other. The novel policy can substantially reduce the temptation to lose in the FIFA World Cup.
\citet{ChaterArrondelGayantLaslier2021} develop a general method to assess the competitiveness of matches played in the last round of the FIFA World Cup group stage. The choice of teams playing each other is found to be crucial to see exciting and fair games. We will adopt their thoroughly tested simulation model in Section~\ref{Sec3}.
\citet{KrumerMegidishSela2023} study the subgame perfect equilibrium of round-robin all-pay tournaments with four symmetric players and two identical prizes. It is shown that one winner in the first round maximises the expected payoff by losing in the second round, but this problem does not occur if there is only one prize \citep{KrumerMegidishSela2017a}.
\citet{CsatoMolontayPinter2024} determine the probability of stakeless matches (when one team or both teams are indifferent since the outcome does not affect their rank) in the UEFA Champions League group stage under alternative schedules.
Finally, \citet{Stronka2024} revisits the issue of tanking in groups of three by demonstrating that two innovative proposals, random tie-breaking (a deliberately biased lottery based on goal differences) and dynamic scheduling, are able to substantially mitigate the risk of collusion.

Our other proposal, reforming the allocation of group labels, has received much less attention because most of the literature takes a within group perspective. Nonetheless, this is an important issue for football tournaments with 24 teams \citep{Guyon2018a}, and will emerge in the 2026 FIFA World Cup, organised with 12 groups of four teams each \citep{FIFA2023b}. Its implementation may modify the kick-off times of certain matches compared to the default setting, which can affect attendance as a share of the capacity of the stadium \citep{Krumer2020a}.

Finally, a radical reform of the knockout bracket can also be powerful against tanking if the teams performing best during the group stage obtain the right to choose their opponents \citep{Guyon2022a, HallLiu2024}. However, adopting this scheme for the FIFA World Cup either lengthens the tournament or requires additional constraints at the expense of fairness \citep{Guyon2022a}. Furthermore, the teams playing the last group matches will enjoy an unfair advantage. Hence, our suggestions seem to have a much higher chance of being implemented in practice.

\section{Methodology} \label{Sec3}

This section presents a framework that allows quantifying the risk of tanking.
The approach outlined in Section~\ref{Sec31} can be applied in any sports competition where a group stage is followed by a knockout phase with pretermined bracket, taking the limitations detailed in Section~\ref{Sec34} into account. Section~\ref{Sec32} customises this model for the FIFA World Cup, and Section~\ref{Sec33} presents some alternatives available for the organiser to mitigate the threat of tanking.

\subsection{An approach to quantify the threat of tanking via simulations} \label{Sec31}

Assume that the top two teams from each group advance to the knockout stage, where the winner of group $i$ plays against the runner-up of the pair of group $i$, denoted by $i^\ast$. The last rounds of these pairs of groups are played at different times. Therefore, if group $i$ is finished, any team in group $i^\ast$ knows its next opponent before the two matches played in the final round if it finishes as the winner (the runner-up of group $i$) or as the runner-up (the winner of group $i$).
This creates an incentive for tanking in group $i^\ast$ if the winner of group $i$ is perceived to be weaker than the runner-up of group $i$.

Suppose that there exists a reliable measure of team strength $\mathcal{S}$, and these ratings are common knowledge.
The risk of tanking emerges in group $i^\ast$ if the winner of group $i$ has a lower value of $\mathcal{S}$ than the runner-up of group $i$.

Thus, a simulation run can consist of the following steps:
\begin{enumerate}
\item
The outcomes of all matches in groups $i$ and $i^\ast$ are generated according to a standard simulation model;
\item
Final rankings in groups $i$ and $i^\ast$ are determined;
\item
Variable $k$ is set as $k=1$;
\item \label{Step4}
If the winner of group $i$ has a lower value of $\mathcal{S}$ than the runner-up of group $i$, then the score of the team that plays against the winner of group $i^\ast$ in the last round is increased by $k$ in this particular match since the winner of group $i^\ast$ attempts using a tanking strategy;
\item
The final ranking in group $i^\ast$ is recomputed because the original group winner conceded $k$ additional goals in its last match, see Step~\ref{Step4};
\item
A tanking opportunity with $k$ goals is registered in group $i^\ast$ if the original winner of group $i^\ast$ is the runner-up under the changed results, implying that a successful tanking strategy exists. \\
Otherwise, $k$ is increased by one and the process returns to Step~\ref{Step4}.
\end{enumerate}
The reason behind the modification of game outcomes is that the winner of group $i^\ast$ will be better off by allowing its opponent to score more goals if it  becomes the runner-up after this manipulation. The simulation run is finished without finding a tanking opportunity if $k > 5$, namely, a successful tanking by the original winner of group $i^\ast$ is impossible or requires that more than 5 additional goals are scored by its opponent in the last round. The definition of collusion by \citet{ChaterArrondelGayantLaslier2021} stops at 5 goals difference, too.

A straightforward measure of tanking $T$ is simply the average probability of tanking opportunities in percentages across all simulation runs.
Analogously, $T$ is defined as the average of $T$ for all pairs of groups if more than one pair of groups is considered, that is, the common tanking measure remains the (average) probability of tanking in the last round of group matches.

However, the cost of a tanking opportunity for the organiser is likely not uniform.
First, the temptation to tank in group $i^\ast$ is stronger if the rating of the winner ($S_1$) minus the rating of the runner-up ($S_2$) in group $i$ is smaller: $S_2 - S_1$ can be regarded as the payoff of a successful tanking. A team will probably strive to tank more resolutely if the potential gain is higher, as the motivational quote at the beginning of the paper illustrates.
Second, tanking is easier to implement if $k$, the number of goals that need to be conceded, is smaller. Hence, it is reasonable to associate a weight for each tanking opportunity:
\[
W = \frac{S_2 - S_1}{100k},
\]
where $100$ is a normalising factor in the denominator (the value of home advantage in the Elo rating).
Consequently, the threat of tanking is more threatening if the runner-up is much stronger than the winner in group $i$, and it can be achieved by letting the opponent score a lower number of additional goals.

The weighted tanking measure $\mathit{WT}$ is the average of weighted tanking opportunities across all simulation runs.
Analogously, $\mathit{WT}$ is defined as the average of $\mathit{WT}$ for all pairs of groups if more than one pair of groups is considered, that is, the common weighted tanking measure is the average of weighted tanking in the last round of group matches.

\subsection{Application for the 2022 FIFA World Cup} \label{Sec32}

The FIFA World Cup takes place every four years and has been organised in the same format between 1998 and 2022. The group stage has been played in eight groups of four teams each, where the top two teams have qualified for the Round of 16. The groups have been paired in alphabetical order: A--B, C--D, E--F, G--H, and the group winner of any group $i$ has played against the runner-up from group $i^\ast$, the pair of group $i$.
 
The limited historical data cannot be used to reliably assess the threat of tanking. However, they provide a solid basis to simulate the outcomes of the games. Thus, an arbitrary number of fictional but reasonable tournaments can be generated as usual in the tournament design literature \citep{LasekGagolewski2018, LeyVandeWieleVanEeetvelde2019, ScarfYusofBilbao2009}.

The prediction model for individual games is a fundamental element of any simulation method. Poisson models, first suggested by \citet{Maher1982}, are perhaps the most popular to generate football match results. According to the underlying assumption, the number of goals scored by both teams follows a Poisson distribution. In particular, team $i$ scores $k$ goals against team $j$ with a probability
\begin{equation*} \label{Poisson_dist}
P_{ij}(k) = \frac{ \left( \lambda_{ij}^{(f)} \right)^k \exp \left( -\lambda_{ij}^{(f)} \right)}{k!},
\end{equation*}
where $\lambda_{ij}^{(f)}$ is the expected number of goals scored in this match played on field $f$.

We adopt the model of \citet{ChaterArrondelGayantLaslier2021}, which is fitted on the basis of all (192) matches played in the first two rounds of the World Cup group stage between 1998 and 2018. The last round of games is not taken into account because other factors than team performance may play a role in determining the result. \citet{ChaterArrondelGayantLaslier2021} have compared this approach to some alternatives and concluded that its parameters can be clearly interpreted, is not worse than other exact-score models, and is competitive with an ordered logistic regression to predict wins, draws, and losses.

Therefore,
\begin{equation} \label{eq_lambda}
\lambda_{ij} = \alpha \frac{R_i}{R_i + R_j},
\end{equation}
where $R_i$ and $R_j$ are the strengths of teams $i$ and $j$, respectively, while parameter $\alpha$ equals the average number of goals scored in the sample.

A team's strength is based on a single variable, its Elo rating. Although FIFA adopted the Elo method of calculation after the 2018 World Cup for its own World Ranking \citep{FIFA2018c}, the formula does not take into account two crucial aspects, home advantage and the margin of victory. Furthermore, the FIFA World Ranking suffered from severe shortcomings before 2018 \citep{CeaDuranGuajardoSureSiebertZamorano2020, Csato2021a, Kaminski2022, LasekSzlavikGagolewskiBhulai2016}. On the other hand, the World Football Elo Ratings seem to be a reliable measure of teams' abilities \citep{GasquezRoyuela2016, HvattumArntzen2010, LasekSzlavikBhulai2013}, and is a widely used benchmark in the literature \citep{ChaterArrondelGayantLaslier2021, Csato2022a, Stronka2020, Stronka2024}.

Since the raw Elo indices fluctuate between 1400 and 2200 in the FIFA World Cup, \citet{ChaterArrondelGayantLaslier2021} adjust the performance differences by a linear transformation of the original Elo rating $E_i$:
\begin{equation} \label{eq_strength}
R_i = 1 + \exp \left( \beta \right) \cdot \frac{E_i - E_{\min}}{E_{\max} - E_{\min}},
\end{equation}
which is finally plugged into~\eqref{eq_lambda}. Here $E_{\min} = 1471$ (the minimal Elo rating in the sample) and $E_{\max} = 2142$ (the maximal Elo rating in the sample).

The model contains two parameters, $\alpha$ in formula~\eqref{eq_lambda} and $\beta$ in formula~\eqref{eq_strength}. The estimations of \citet[Table~5]{ChaterArrondelGayantLaslier2021} are adopted, namely, $\alpha = 2.5156$, which equals the average number of goals per match in the sample, and $\beta = 3.7581$.
\citet{ChaterArrondelGayantLaslier2021} calculate a number of statistics from the chosen Poisson model in order to compare them with the sample data. According to these tests, the simulation framework provides a good approximation of the actual score-generating process.

\begin{table}[t!]
  \centering
  \caption{Groups and teams in the 2022 FIFA World Cup}
  \label{Table1}
    \rowcolors{1}{gray!20}{}
    \begin{tabularx}{0.9\textwidth}{LCCLC} \toprule \hiderowcolors
    Country & Elo &       & Country & Elo \\ \midrule
    \multicolumn{2}{c}{\textbf{Group A}} &       & \multicolumn{2}{c}{\textbf{Group B}} \\ \bottomrule \showrowcolors
    Qatar & 1663  &       & England & 2039 \\
    Netherlands & 1938  &       & United States & 1822 \\
    Senegal & 1729  &       & Iran  & 1820 \\
    Ecuador & 1840  &       & Wales & 1841 \\ \toprule
    \end{tabularx}
    
    \rowcolors{1}{}{gray!20}
    \begin{tabularx}{0.9\textwidth}{LCCLC} \toprule \hiderowcolors
    \multicolumn{2}{c}{\textbf{Group C}} &       & \multicolumn{2}{c}{\textbf{Group D}} \\ \bottomrule \showrowcolors
    Argentina & 2108  &       & France & 2116 \\
    Mexico & 1848  &       & Denmark & 1936 \\
    Poland & 1799  &       & Australia & 1677 \\
    Saudi Arabia & 1634  &       & Tunisia & 1612 \\ \toprule
    \end{tabularx}
    
    \rowcolors{1}{}{gray!20}
    \begin{tabularx}{0.9\textwidth}{LCCLC} \toprule \hiderowcolors
    \multicolumn{2}{c}{\textbf{Group E}} &       & \multicolumn{2}{c}{\textbf{Group F}} \\ \bottomrule \showrowcolors
    Spain & 2039  &       & Belgium & 2069 \\
    Germany & 1966  &       & Croatia & 1855 \\
    Japan & 1796  &       & Morocco & 1738 \\
    Costa Rica & 1743  &       & Canada & 1798 \\ \toprule
    \end{tabularx}

\begin{threeparttable}    
    \rowcolors{1}{}{gray!20}
    \begin{tabularx}{0.9\textwidth}{LCCLC} \toprule \hiderowcolors
    \multicolumn{2}{c}{\textbf{Group G}} &       & \multicolumn{2}{c}{\textbf{Group H}} \\ \bottomrule \showrowcolors
    Brazil & 2155  &       & Portugal & 1984 \\
    Switzerland & 1920  &       & Uruguay & 1923 \\
    Serbia & 1845  &       & South Korea & 1800 \\
    Cameroon & 1631  &       & Ghana & 1541 \\ \toprule
    \end{tabularx}
\begin{tablenotes} \footnotesize
\item
The rating of the host Qatar is increased by 100, the fixed value of home advantage in the World Football Elo Ratings, see \url{http://eloratings.net/about}.
\item
The column Elo shows the strength of the teams according to the World Football Elo Ratings as of 31 March 2022, see \url{https://www.international-football.net/elo-ratings-table?year=2022&month=03&day=31}.
\end{tablenotes}
\end{threeparttable}
\end{table}

Finally, the strengths of the teams playing in the 2022 FIFA World Cup should be determined. Elo rating is a dynamic measure, updated for both teams after they play a match. Consequently, Elo ratings move between the draw and the beginning of the tournament, as well as during the tournament. Since we think the appropriate time to determine the schedule of the games coincides with the draw of the groups that took place on 1 April 2022, the Elo ratings of 31 March 2022 are used to measure team abilities, see Table~\ref{Table1}.\footnote{~\citet{Csato2023d} used the same Elo ratings to analyse the 2022 FIFA World Cup draw but the rating of Argentina contained a typo: it was mistakenly considered to be 2018 instead of the correct 2108.}

The process outlined in Section~\ref{Sec31} requires to rank the teams on the basis of group matches.
The ranking criteria follow the official rule \citep[Article~12]{FIFA2022c}:
(1) higher number of points obtained in all group matches;
(2) superior goal difference in all group matches;
(3) higher number of goals scored in all group matches;
(4) higher number of points obtained in the group matches between the teams concerned;
(5) superior goal difference resulting from the group matches between the teams concerned;
(6) higher number of goals scored in the group matches between the teams concerned;
(7) drawing of lots.
The fair play points based on yellow and red cards are not considered despite this criterion has been applied before drawing of lots in the FIFA World Cup.

\begin{table}[t!]
  \centering
  \caption{Matches played in the last round of the 2022 FIFA World Cup}
  \label{Table2}
\begin{threeparttable}
    \rowcolors{1}{}{gray!20}
    \begin{tabularx}{0.9\textwidth}{lLLc} \toprule
    Group & Team 1 & Team 2 & Type of schedule \\ \bottomrule
    Group A & Qatar (Pot 1) & Netherlands (Pot 2) & Schedule A \\
          & Senegal (Pot 3) & Ecuador (Pot 4) & \\ \hline
    Group B & England (Pot 1) & Wales (Pot 4) & Schedule C \\
          & United States (Pot 2) & Iran (Pot 3) & \\ \hline
    Group C & Argentina (Pot 1) & Poland (Pot 3) & Schedule B \\
          & Mexico (Pot 2) & Saudi Arabia (Pot 4) & \\ \hline
    Group D & France (Pot 1) & Tunisia (Pot 4) & Schedule C \\
          & Denmark (Pot 2) & Australia (Pot 3) & \\ \hline
    Group E & Spain (Pot 1) & Japan (Pot 3) & Schedule B \\
          & Germany (Pot 2) & Costa Rica (Pot 4) & \\ \hline
    Group F & Belgium (Pot 1) & Croatia (Pot 2) & Schedule A \\
          & Morocco (Pot 3) & Canada (Pot 4) & \\ \hline
    Group G & Brazil (Pot 1) & Cameroon (Pot 4) & Schedule C \\
          & Switzerland (Pot 2) & Serbia (Pot 3) & \\ \hline
    Group H & Portugal (Pot 1) & South Korea (Pot 3) & Schedule B \\
          & Uruguay (Pot 2) & Ghana (Pot 4) & \\ \bottomrule
    \end{tabularx}
\begin{tablenotes} \footnotesize
\item
The Pot from which the team was drawn is shown in parenthesis.
\end{tablenotes}
\end{threeparttable}
\end{table}

The matches played in the last round of the group stage, used to determine the team against which the tanking strategy is applied, are presented in Table~\ref{Table2}.
Furthermore, since the threat of tanking for any pair of groups primarily emerges in the group that is finished later, the order of the kick-off times in the last round should also be taken into account. In the 2022 FIFA World Cup, the scheme A--B, D--C, F--E, H--G has been followed, that is, the order of the groups has been determined randomly for any pair. This is reinforced by the previous two editions of the tournament: B--A, D--C, F--E, G--H (2014 FIFA World Cup) and A--B, C--D, F--E, H--G (2018 FIFA World Cup). 
To summarise, the final rounds in the four pairs of groups have been played on subsequent days, but one group has always finished before the beginning of the last two matches in its pair---simultaneously playing all the four games seems to be infeasible because of commercial reasons.

The simulation process is repeated 1 million times.

\subsection{Decision variables} \label{Sec33}

The simulation model will be applied to find an optimal schedule. In particular, the group composition of the 2022 FIFA World Cup is assumed to be given, see Table~\ref{Table1}. However, the organiser has two available options immediately after the group draw in order to influence the threat of tanking: choosing the group labels and deciding the schedule of group matches.

Three different group labelling rules are considered:
\begin{itemize}
\item
\emph{Random}: the groups are labelled randomly after the draw, which is equivalent to the lack of relabelling;
\item
\emph{Pair optimal}: for each pair of groups, the kick-off times in the last round are chosen separately to minimise the threat of tanking;
\item
\emph{Optimal}: the groups are relabelled after the draw to minimise the threat of tanking.
\end{itemize}

For example, measure $T$ suggests that it would have been better to finish Group A later than Group B, in contrast to the schedule of the 2022 FIFA World Cup. This could have been achieved immediately after the draw, by exchanging the kick-off times of the matches played in the last round, without altering their day.
On the other hand, the optimal solution contains the pair F--A, namely, Group F should have been labelled Group B and finished before Group A in order to minimise the threat of tanking. Although this requires a more radical intervention, it remains feasible directly after the draw. Naturally, the optimal solution may depend on the measure of tanking $T$ or $\mathit{WT}$.

Since there are four pairs of groups, the set of alternatives available under the pair optimal group labelling rule is $2^4 = 16$. In the case of the optimal rule, there are seven options for the first pair of groups, five options for the second pair, and three options for the third pair. However, the order can be chosen independently for each pair, hence, the optimal solution should be found among $7 \times 5 \times 3 \times 16 = 1680$ scenarios.

Analogously, three different group match scheduling policies are examined:
\begin{itemize}
\item
\emph{Random}: the schedule of the games is decided randomly, independently for each group, which is the current policy of FIFA \citep{ChaterArrondelGayantLaslier2021};
\item
\emph{Optimal uniform}: the same schedule of the games is chosen for all groups based on the seeding of the teams to minimise the risk of tanking;
\item
\emph{Optimal}: the schedule of the games is determined in each group separately to minimise the risk of tanking.
\end{itemize}

Each group contains one team from each of Pots 1--4 \citep{Csato2023d}.
Therefore, the uniform schedule allows for three options in the last round of group matches:
\begin{itemize}
\item
\emph{Schedule A}: Pot 1 vs.\ Pot 2 and Pot 3 vs.\ Pot 4;
\item
\emph{Schedule B}: Pot 1 vs.\ Pot 3 and Pot 2 vs.\ Pot 4;
\item
\emph{Schedule C}: Pot 1 vs.\ Pot 4 and Pot 2 vs.\ Pot 3.
\end{itemize}
The last column of Table~\ref{Table2} shows the randomly drawn schedules in the 2022 FIFA World Cup.
Consequently, the optimal scheduling policy needs to assess $3^8 = 6561$ possibilities.

From the viewpoint of logistics, it seems entirely sufficient to determine the labels of the groups and the schedule of group matches on the day of the group draw, which allows running some simulations after the composition of the groups is already known.

\subsection{Limitations} \label{Sec34}

The simulation model, adopted from \citet{ChaterArrondelGayantLaslier2021}, is not necessarily the best available option even for the FIFA World Cup.
However, the exact prediction method is not an essential part of our approach. The tournament designer is free to choose any exact-score generating model according to its preferences, repeat the process described in Section~\ref{Sec31}, and choose an optimal schedule from the set of alternatives given in Section~\ref{Sec33}.

Besides the standard caveats of using simulation models, the identification of tanking remains imperfect because of the following reasons:
\begin{itemize}
\item
The teams do not focus only on their opponent in the Round of 16 but on the difficulty of the whole knockout bracket (e.g.\ a team may intend to avoid playing the favourite team before the final even if it implies a stronger opponent in the Round of 16);
\item
Elo rating is not necessarily a good indicator of preferences in the Round of 16 (e.g.\ a European team might want to avoid playing against another European team in the knockout stage);
\item
Tanking is assumed to be a unilateral action from the original winner of group $i^\ast$, however, the other competitors for the top two positions in group $i^\ast$ can also take the possible manipulation of this team into account (when a complex interaction between two or more teams that are all interested in tanking should be modelled);
\item
Tanking strategies may be applied even if there is uncertainty in the opponents in the Round of 16 (e.g.\ a team may be tempted to avoid a strong team even if it is only a likely runner-up);
\item
The group winner might be known after two rounds are played (but, according to \citet[p.~9]{Stronka2020}, such a situation has a negligible probability);
\item
A team may be not willing to lose even if this strategy maximises the expected payoff;
\item
Schedule effects are neglected, which is probably not a problem with respect to carry-over \citep{GoossensSpieksma2012a} but there exists empirical evidence that the team playing the first matches in the first two rounds has a higher probability of qualifying for the next stage in the FIFA World Cup \citep{KrumerLechner2017};
\item
The simulation model does not take momentum into account, which is an important factor of performance according to the beliefs of athletes and sports fans, even though the academic literature usually finds it a fallacy \citep{Vergin2000, SteegerDulinGonzalez2021}.
\end{itemize}

In our opinion, these compromises are necessary to get a tractable model of tanking that remains understandable for sports governing bodies.
In addition, even though the probability of tanking derived from the simulations may be inaccurate and has limited meaning on its own due to the above shortcomings, the results can be appropriate to \emph{compare} alternative tournament designs \citep{Appleton1995}.

\section{Numerical evaluation of tournament designs for the 2022 FIFA World Cup} \label{Sec4}

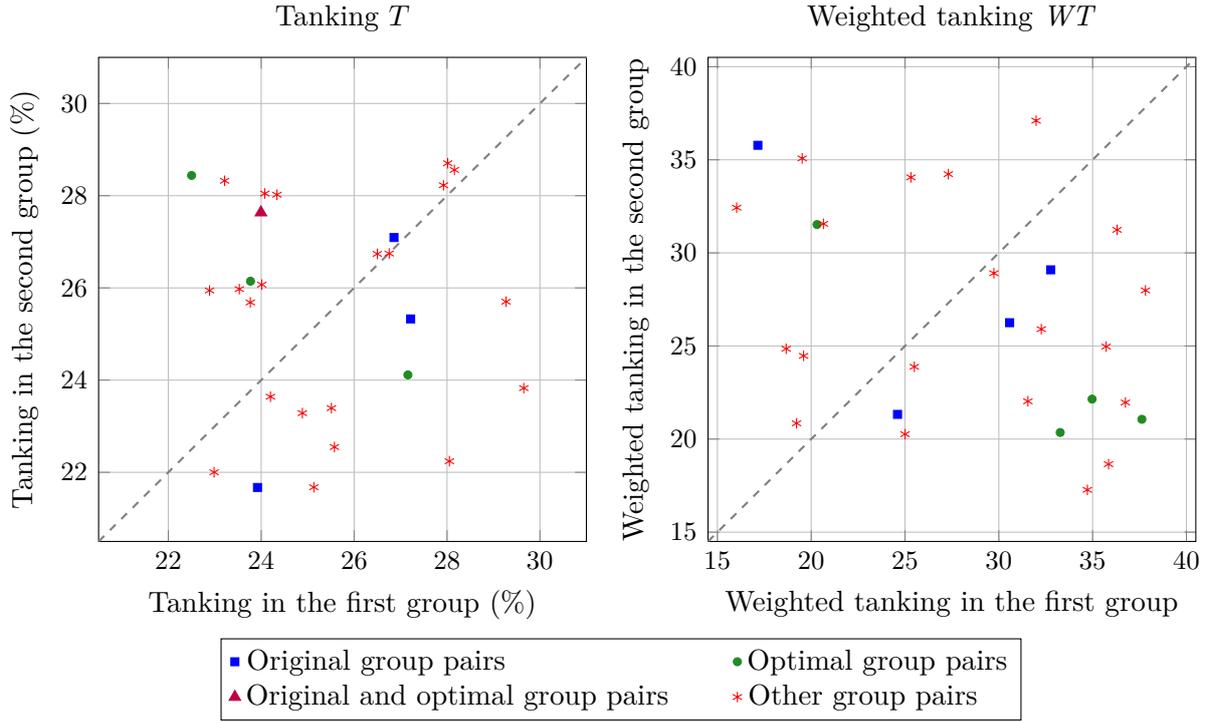
\begin{figure}[t!]
\centering

\begin{tikzpicture}
\begin{axis}[
name = axis1,
title = {Tanking $T$},
title style = {font=\small},
xlabel = Tanking in the first group (\%),
x label style = {font=\small},
ylabel = Tanking in the second group (\%),
y label style = {font=\small},
width = 0.5\textwidth,
height = 0.5\textwidth,
nodes near coords,
xmajorgrids = true,
ymajorgrids = true,
xmin = 20.5,
xmax = 31,
ymin = 20.5,
ymax = 31,
legend style = {font=\small,at={(0.25,-0.2)},anchor=north west,legend columns=2},
legend entries = {Original group pairs$\qquad \qquad \qquad \quad \; \;$, Optimal group pairs, Original and optimal group pairs$\qquad$, Other group pairs$\quad$},
]
\addplot [scatter,blue,only marks,mark size=1.5pt,mark=square*,point meta=explicit symbolic]
coordinates {
(26.8584,27.0924)
(23.9197,21.6716)
(27.2167,25.3271)
};
\addplot [scatter,ForestGreen,only marks,mark size=1.5pt,point meta=explicit symbolic]
coordinates {
(23.7698,26.145)
(22.5013,28.4394)
(27.1578,24.1126)
};
\addplot [scatter,purple,only marks,mark size=2.5pt,mark=triangle*,point meta=explicit symbolic]
coordinates {
(23.994,27.6346)
};
\addplot [scatter,red,only marks,mark size=2pt,mark=asterisk,point meta=explicit symbolic,nodes near coords style = {text=red, anchor=south east}]
coordinates {
(23.2127,28.3234)
(23.5292,25.9735)
(26.4993,26.7373)
(24.0779,28.047)
(28.0147,28.7042)
(22.8876,25.9435)
(26.7549,26.7455)
(24.3388,28.0214)
(24.0122,26.0711)
(28.1584,28.5593)
(28.0512,22.2403)
(25.5116,23.3921)
(25.133,21.6754)
(29.652,23.8263)
(25.5765,22.5485)
(24.1997,23.64)
(22.9864,22.0007)
(23.7664,25.6843)
(27.9242,28.2263)
(24.8863,23.2837)
(29.2707,25.7011)
};
\draw [gray,dashed,thick] (rel axis cs:0,0) -- (rel axis cs:1,1);
\end{axis}

\begin{axis}[
at = {(axis1.south east)},
xshift = 0.1\textwidth,
title = {Weighted tanking $\mathit{WT}$},
title style = {font=\small},
xlabel = Weighted tanking in the first group,
x label style = {font=\small},
ylabel = Weighted tanking in the second group,
y label style = {font=\small},
width = 0.5\textwidth,
height = 0.5\textwidth,
nodes near coords,
xmajorgrids = true,
ymajorgrids = true,
xmin = 14.5,
xmax = 40.5,
ymin = 14.5,
ymax = 40.5,
]
\addplot [scatter,blue,only marks,mark size=1.5pt,mark=square*,point meta=explicit symbolic]
coordinates {
(24.6007706500168,21.3260698166487)
(32.767498199978,29.0902448999779)
(30.5760889833874,26.2507453833611)
(17.1556141666429,35.7792367665972)
};
\addplot [scatter,ForestGreen,only marks,mark size=1.5pt,point meta=explicit symbolic]
coordinates {
(37.6242312499654,21.0616255499808)
(33.2729423833092,20.3517959000081)
(34.9713702332714,22.1464162166839)
(20.3110037832955,31.5224696667226)
};
\addplot [scatter,red,only marks,mark size=2pt,mark=asterisk,point meta=explicit symbolic,nodes near coords style = {text=red, anchor=south east}] coordinates {
(34.7195524666418,17.2720502166498)
(25.0028922000254,20.2630728666483)
(31.5507961167219,22.0287324666474)
(35.8498327499327,18.6529845666494)
(19.21914912,20.84493333)
(35.7211882999706,24.9686424666818)
(25.4969557333603,23.882851916682)
(32.2642352000562,25.9030473833521)
(36.7463046999271,21.9637866000116)
(19.5938472332978,24.4692895833477)
(25.3145762833602,34.0553975999729)
(31.988089250056,37.1013388499663)
(36.3032622999281,31.2375878166443)
(19.5203035666336,35.0756534499724)
(20.6522843333465,31.5609086166441)
(27.3074289833859,34.2292349499751)
(29.7384081666291,28.906022783314)
(16.026464816648,32.4281039999791)
(18.6669169666361,24.8509814666925)
(37.8155348999195,27.9847201000518)
};
\draw [gray,dashed,thick] (rel axis cs:0,0) -- (rel axis cs:1,1);
\end{axis}
\end{tikzpicture}

\caption{Measures of tanking for the 28 possible group pairs in the \\ 2022 FIFA World Cup, original schedule in each group (see Table~\ref{Table2})}
\label{Fig1}

\end{figure}


Figure~\ref{Fig1} shows the probability of tanking for each possible pair of the eight groups listed in Table~\ref{Table1} if the schedule used in the 2022 FIFA World Cup is followed in all groups. A point over (below) the dashed 45 degree line indicates that the group coming first (second) in alphabetical order should be finished later than its pair to mitigate the threat of tanking. The optimal solution, represented by the green dots (and the purple triangle), is G--A, C--B, D--H, F--E according to measure $T$, and A--C, B--D, E--G, H--F according to measure $\mathit{WT}$.
Thus, weighted tanking usually decreases when groups with an outstanding team are finished later in order to avoid the strong temptation of tanking in the other group if the strongest team finishes only in the second place due to bad luck. Groups C, D, G exhibit this pattern as can be seen in Table~\ref{Table1}.
While the optimal schedule is quite sensitive to the tanking measure, this can be chosen according to the preferences of the decision-maker.

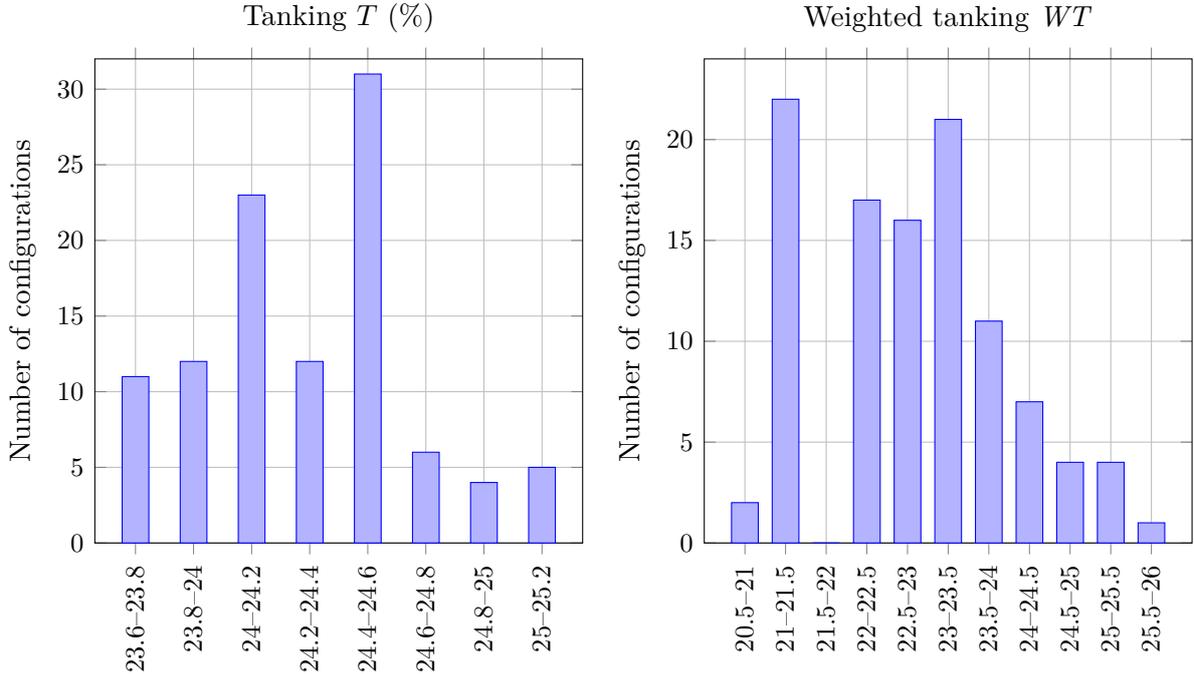
\begin{figure}[t!]
\centering

\begin{tikzpicture}
\begin{axis}[
name = axis1,
title = {Tanking $T$ (\%)},
title style = {font=\small},
ylabel = {Number of configurations},
ylabel style = {align=center, font=\small},
width = 0.5\textwidth,
height = 0.5\textwidth,
xmajorgrids,
ymajorgrids,
symbolic x coords = {23.6--23.8,23.8--24,24--24.2,24.2--24.4,24.4--24.6,24.6--24.8,24.8--25,25--25.2},
xtick=data,
enlarge x limits = 0.1,
x tick label style = {rotate=90},
ymin = 0,
ymax = 32,
scaled y ticks = false,
ybar,
yticklabel style = {/pgf/number format/fixed,/pgf/number format/precision=5},
]
\addplot coordinates{
(23.6--23.8,11)
(23.8--24,12)
(24--24.2,23)
(24.2--24.4,12)
(24.4--24.6,31)
(24.6--24.8,6)
(24.8--25,4)
(25--25.2,5)
};
\end{axis}

\begin{axis}[
at = {(axis1.south east)},
xshift = 0.1\textwidth,
title = {Weighted tanking $\mathit{WT}$},
title style = {font=\small},
ylabel = {Number of configurations},
y label style = {font=\small},
width = 0.5\textwidth,
height = 0.5\textwidth,
xmajorgrids,
ymajorgrids,
symbolic x coords = {20.5--21,21--21.5,21.5--22,22--22.5,22.5--23,23--23.5,23.5--24,24--24.5,24.5--25,25--25.5,25.5--26,},
xtick=data,
enlarge x limits = 0.1,
x tick label style = {rotate=90},
ymin = 0,
ymax = 24,
scaled y ticks = false,
ybar,
yticklabel style = {/pgf/number format/fixed,/pgf/number format/precision=5},
]
\addplot coordinates{
(20.5--21,2)
(21--21.5,22)
(21.5--22,0)
(22--22.5,17)
(22.5--23,16)
(23--23.5,21)
(23.5--24,11)
(24--24.5,7)
(24.5--25,4)
(25--25.5,4)
(25.5--26,1)
};
\end{axis}
\end{tikzpicture}

\caption{Distribution of measures of tanking for the 105 possible \\ pair optimal group label configurations in the 2022 FIFA World Cup}
\label{Fig2}

\end{figure}


Figure~\ref{Fig2} compares the pair optimal and optimal group labelling rules by showing the distribution of the two measures of tanking for the 105 ($= 7 \times 5 \times 3$) possible allocations of group labels. The impact of relabelling all groups is moderated compared to choosing which group in a given pair should be finished earlier. Regarding the probability of tanking (weighted tanking $\mathit{WT}$), the more favourable group has a relative advantage of 9.2\% (27.6\%) over its pair on average across the 28 pairs of groups as can be seen in Figure~\ref{Fig1}. On the other hand, even the best configuration of the eight groups is able to reduce tanking (weighted tanking) by at most 5.8\% (19.4\%) compared to the worst option, that is, the relative standard deviation remains small. This might be a positive message for the organiser: optimising the kick-off times for the group pairs has a small cost but it ensures a high proportion of the potential gain in mitigating the risk of tanking.

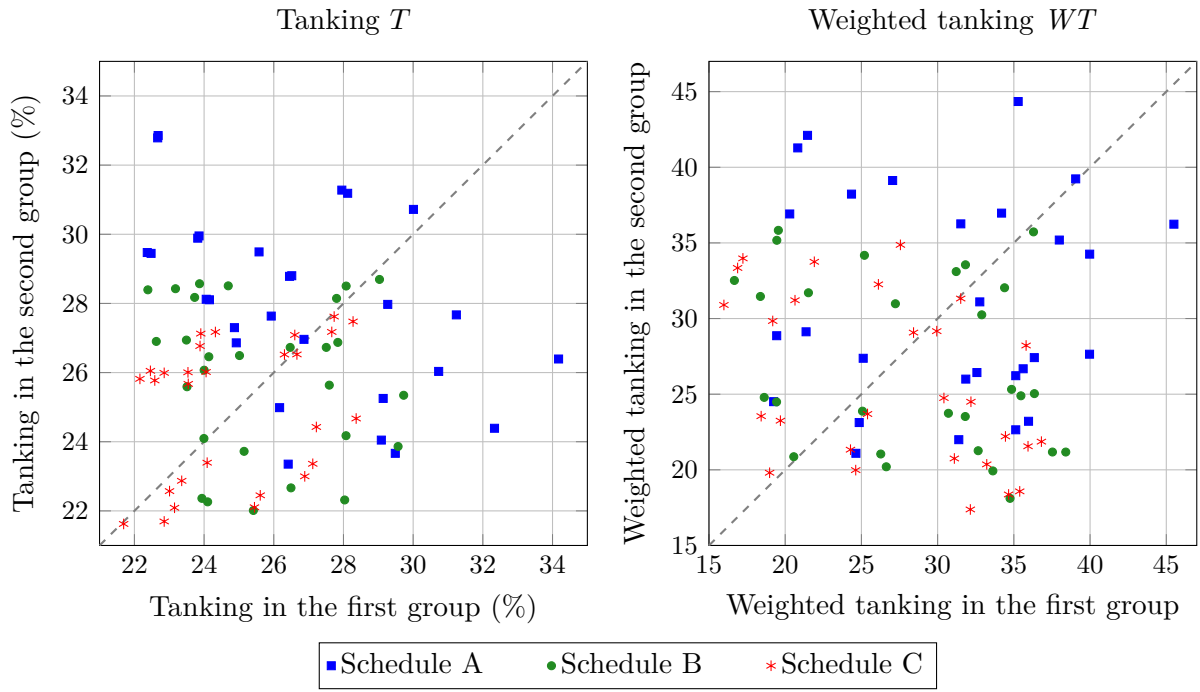
\begin{figure}[t!]
\centering

\begin{tikzpicture}
\begin{axis}[
name = axis1,
title = {Tanking $T$},
title style = {font=\small},
xlabel = Tanking in the first group (\%),
x label style = {font=\small},
ylabel = Tanking in the second group (\%),
y label style = {font=\small},
width = 0.5\textwidth,
height = 0.5\textwidth,
nodes near coords,
xmajorgrids = true,
ymajorgrids = true,
xmin = 21,
xmax = 35,
ymin = 21,
ymax = 35,
]
\addplot [scatter,blue,only marks,mark size=1.5pt,mark=square*,point meta=explicit symbolic]
coordinates {
(26.8666,26.9611)
(22.3692,29.4718)
(22.6806,32.8564)
(26.452,28.7834)
(24.0613,28.1205)
(23.818,29.8868)
(27.9529,31.2771)
(22.4784,29.4473)
(22.6683,32.7876)
(26.5155,28.8009)
(24.1534,28.1051)
(23.8554,29.952)
(28.1202,31.1854)
(24.871,27.2995)
(29.0869,24.0476)
(26.4137,23.3521)
(26.1671,24.9854)
(30.7322,26.0337)
(32.3347,24.3895)
(29.4877,23.665)
(29.1395,25.2524)
(34.1737,26.3928)
(25.9271,27.634)
(25.5772,29.4897)
(30.009,30.7194)
(24.9239,26.8599)
(29.2711,27.9723)
(31.2399,27.6685)
};
\addplot [scatter,ForestGreen,only marks,mark size=1.5pt,point meta=explicit symbolic]
coordinates {
(27.8378,26.873)
(23.1793,28.4255)
(23.4926,26.94)
(27.5065,26.7284)
(25.0164,26.4919)
(24.6933,28.5085)
(29.0342,28.6938)
(22.3864,28.3932)
(22.6283,26.9012)
(26.4707,26.7278)
(24.1356,26.458)
(23.8698,28.5696)
(28.0768,28.5027)
(23.9375,22.3623)
(28.035,22.3161)
(25.4142,22.0122)
(25.1481,23.7222)
(29.5677,23.8629)
(26.4936,22.6657)
(24.102,22.2615)
(23.9954,24.0943)
(28.0734,24.1747)
(24.0018,26.0692)
(23.7288,28.1742)
(27.8022,28.1454)
(23.5097,25.5899)
(27.5933,25.6364)
(29.7278,25.3438)
};
\addplot [scatter,red,only marks,mark size=2pt,mark=asterisk,point meta=explicit symbolic,nodes near coords style = {text=red, anchor=south east}]
coordinates {
(26.603,27.0873)
(22.1585,25.8184)
(22.4594,26.0498)
(26.3066,26.5212)
(23.9056,27.1272)
(23.5373,26.0073)
(27.7331,27.6198)
(22.5848,25.7722)
(22.855,25.9917)
(26.6642,26.5252)
(24.3246,27.1736)
(24.0573,26.0168)
(28.2755,27.4773)
(21.6912,21.6223)
(25.4459,22.1048)
(23.0083,22.5718)
(22.855,21.6944)
(26.8892,22.9996)
(25.6117,22.4502)
(23.3577,22.8694)
(23.1516,22.0901)
(27.1193,23.3621)
(23.8837,26.7683)
(23.5458,25.6726)
(27.6646,27.1756)
(24.0883,23.3942)
(28.3644,24.6716)
(27.222,24.427)
};
\draw [gray,dashed,thick] (rel axis cs:0,0) -- (rel axis cs:1,1);
\end{axis}

\begin{axis}[
at = {(axis1.south east)},
xshift = 0.1\textwidth,
title = {Weighted tanking $\mathit{WT}$},
title style = {font=\small},
xlabel = Weighted tanking in the first group,
x label style = {font=\small},
ylabel = Weighted tanking in the second group,
y label style = {font=\small},
width = 0.5\textwidth,
height = 0.5\textwidth,
nodes near coords,
xmajorgrids = true,
ymajorgrids = true,
xmin = 15,
xmax = 47,
ymin = 15,
ymax = 47,
legend style = {font=\small,at={(-0.8,-0.2)},anchor=north west,legend columns=3},
legend entries = {Schedule A$\qquad$, Schedule B$\qquad$, Schedule C},
]
\addplot [scatter,blue,only marks,mark size=1.5pt,mark=square*,point meta=explicit symbolic]
coordinates {
(24.6391755333515,21.0929496999823)
(35.1233188,22.64588867)
(32.5769175999762,26.4305090333111)
(24.8556787666912,23.1245567833152)
(31.3880545833887,21.9923219999813)
(35.9631307332652,23.2099339833135)
(19.2436603166332,24.5064893333126)
(35.610511416637,26.6798870500155)
(32.7657533833095,31.1022730499931)
(25.1177053000253,27.3741088666808)
(31.84756158,25.99717265)
(36.3417831165966,27.4171711500121)
(19.4431973999645,28.86525261667)
(35.2832163666439,44.3468449832835)
(27.0486969500265,39.1249720666273)
(34.1945518167251,36.9695941666331)
(39.056804066578,39.2332630332963)
(20.8250848832937,41.2800830499581)
(31.5203500333421,36.2597128833081)
(39.9733197334015,34.2575864499751)
(45.4991947998695,36.2376892333094)
(24.3503834332781,38.2263882666425)
(35.1261380333941,26.221612583362)
(39.96512523324,27.6434122333571)
(21.3699342832902,29.1243213500186)
(37.982479733252,35.189798016727)
(20.2969199166285,36.914176100063)
(21.464989666625,42.1109057665583)
};
\addplot [scatter,ForestGreen,only marks,mark size=1.5pt,point meta=explicit symbolic]
coordinates {
(26.2691854833536,21.0453305166494)
(37.53212315,21.17705223)
(34.7492350999746,18.1141358833149)
(26.6312912666942,20.2022388499825)
(33.6228178500589,19.9282261999821)
(38.4053202832503,21.1760257999802)
(20.5631194999604,20.8672162166471)
(35.4627045166369,24.893675266682)
(32.6592460166421,21.2636817000079)
(25.0705349166922,23.8783758166824)
(31.8124869,23.52489147)
(36.3482362165963,25.0377577500152)
(19.4341370832979,24.4892611166803)
(32.8946390499782,30.2476002833112)
(25.1979613666936,34.1796713499736)
(31.8270288500557,33.5549876333075)
(36.289614066595,35.7283924666363)
(19.4504680833002,35.1731135333047)
(21.5246691500163,31.7050456499774)
(27.2301356667187,30.9787384499772)
(31.2182731499546,33.1055047333118)
(16.667385016646,32.5178241666455)
(30.707119750054,23.7375065833547)
(34.8507036832716,25.3195155666936)
(18.6244518166363,24.7881478500252)
(34.3905892499412,32.0355302167217)
(18.3744256999705,31.4604586167219)
(19.5510822833002,35.831267433264)
};
\addplot [scatter,red,only marks,mark size=2pt,mark=asterisk,point meta=explicit symbolic,nodes near coords style = {text=red, anchor=south east}] coordinates {
(24.2836011000171,21.3319268333154)
(34.65574938,18.37080513)
(32.1463332166437,17.3798327499828)
(24.626824966691,19.9820793166494)
(31.0975006500551,20.7497588833151)
(35.3857768332689,18.5760864333163)
(18.9782453499673,19.8067208833149)
(35.929999233303,21.5581705166774)
(33.222564133309,20.3610403333412)
(25.4017409500268,23.6912217166817)
(32.18474138,24.49918182)
(36.8081928165927,21.8636925833446)
(19.6867364166304,23.2487922000122)
(28.4193615499808,29.0727883166453)
(21.9112619000166,33.752202499974)
(27.5593241500507,34.8708419666377)
(31.5146565166229,31.3223715999782)
(16.8793464166435,33.3442121333078)
(20.6436885333469,31.211307283311)
(26.1149576500517,32.2595575499764)
(29.9425857832948,29.1676808833142)
(15.979788233315,30.8982678499792)
(30.4130903333871,24.7480582166918)
(34.4488639832741,22.2096351000177)
(18.4234181833035,23.5479614333549)
(35.8038258499329,28.2188690500522)
(19.1830320332998,29.840036616721)
(17.2266936999757,33.9770731332745)
};
\draw [gray,dashed,thick] (rel axis cs:0,0) -- (rel axis cs:1,1);
\end{axis}
\end{tikzpicture}

\caption{Measures of tanking for the 28 possible group pairs in \\ the 2022 FIFA World Cup as a function of group match schedule}
\label{Fig3}

\end{figure}


Figure~\ref{Fig3} presents how the schedule affects the measures of tanking. According to our simulation model, the results of the matches do not depend on their sequence, thus, only the order of the games in the group that is finished later counts. The probability of tanking is slightly determined by the schedule. Nonetheless, Schedule B is the best choice for 6 (6), and Schedule C is the best for 21 (22) group pairs with respect to measure $T$ ($\mathit{WT}$). Consequently, Schedule A---when the teams from the two strongest pots play against each other in the last round---is almost always dominated.

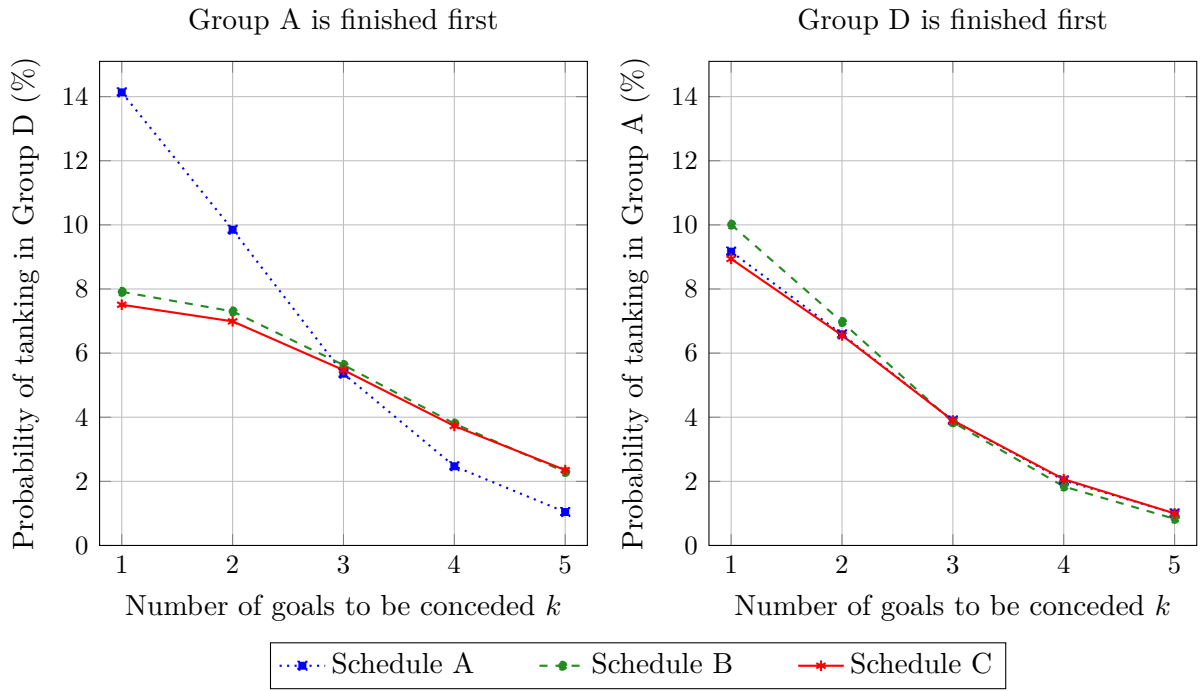
\begin{figure}[t!]
\centering

\begin{tikzpicture}
\begin{axis}[
name = axis1,
title = {Group A is finished first},
title style = {font=\small},
xlabel = Number of goals to be conceded $k$,
x label style = {font=\small},
ylabel = Probability of tanking in Group D (\%),
y label style = {font=\small},
width = 0.5\textwidth,
height = 0.5\textwidth,
nodes near coords,
xmajorgrids = true,
ymajorgrids = true,
xmin = 0.8,
xmax = 5.2,
ymin = 0,
ymax = 15.1,
]
\addplot [scatter,blue,dotted,thick,mark size=1.5pt,mark=square*,point meta=explicit symbolic] coordinates {
(1,14.1342)
(2,9.8495)
(3,5.3579)
(4,2.4716)
(5,1.0432)
};
\addplot [scatter,ForestGreen,dashed,thick,mark size=1.5pt,point meta=explicit symbolic] coordinates {
(1,7.9114)
(2,7.3018)
(3,5.6303)
(4,3.8032)
(5,2.2933)
};
\addplot [scatter,red,thick,mark size=2pt,mark=asterisk,point meta=explicit symbolic,nodes near coords style = {text=red, anchor=south east}] coordinates {
(1,7.5085)
(2,6.9906)
(3,5.4713)
(4,3.7282)
(5,2.3512)
};
\end{axis}

\begin{axis}[
at = {(axis1.south east)},
xshift = 0.1\textwidth,
title = {Group D is finished first},
title style = {font=\small},
xlabel = Number of goals to be conceded $k$,
x label style = {font=\small},
ylabel = Probability of tanking in Group A (\%),
y label style = {font=\small},
width = 0.5\textwidth,
height = 0.5\textwidth,
nodes near coords,
xmajorgrids = true,
ymajorgrids = true,
xmin = 0.8,
xmax = 5.2,
ymin = 0,
ymax = 15.1,
legend style = {font=\small,at={(-0.9,-0.2)},anchor=north west,legend columns=3},
legend entries = {Schedule A$\qquad$, Schedule B$\qquad$, Schedule C},
]
\addplot [scatter,blue,dotted,thick,mark size=1.5pt,mark=square*,point meta=explicit symbolic] coordinates {
(1,9.1661)
(2,6.585)
(3,3.8983)
(4,2.0323)
(5,0.9989)
};
\addplot [scatter,ForestGreen,dashed,thick,mark size=1.5pt,point meta=explicit symbolic] coordinates {
(1,10.0091)
(2,6.9703)
(3,3.8439)
(4,1.8399)
(5,0.8294)
};
\addplot [scatter,red,thick,mark size=2pt,mark=asterisk,point meta=explicit symbolic,nodes near coords style = {text=red, anchor=south east}] coordinates {
(1,8.9399)
(2,6.5542)
(3,3.899)
(4,2.0669)
(5,0.9994)
};
\end{axis}
\end{tikzpicture}

\caption{Probability of tanking by the number of goals \\ conceded for group pair A--D as a function of the schedule}
\label{Fig4}

\end{figure}


Figure~\ref{Fig4} illustrates that the probability of tanking depends to some extent on the number of goals considered. If Groups A and D are paired, group A should be finished first with Schedule C if $k$ is at most one or two, but it becomes optimal to finish first Group D under Schedule A or C if higher values of $k$ are also taken into account. Unsurprisingly, weighted tanking $\mathit{WT}$ suggests playing the last round of Group A earlier since the difference between the Elo ratings of the top teams in Group D is large, thus, there will be a strong temptation to tank in Group A.

\begin{table}[t!]
  \centering
  \caption{Measures of tanking for the 2022 FIFA World Cup groups}
  \label{Table3}
  
\begin{subtable}{\textwidth}
  \centering
  \caption{Probability of tanking $T$ (\%)}
  \label{Table3a}
  \rowcolors{1}{gray!20}{}
    \begin{tabularx}{0.9\textwidth}{lCCcC} \toprule \hiderowcolors
    \multirow{2}{*}{Group labelling rule} & \multicolumn{4}{c}{Group match scheduling policy} \\
    		 & Original & Random & Optimal uniform & Optimal \\ \bottomrule \showrowcolors
    Original & 25.556 & 25.997 & 24.913 & 24.799 \\
    Random & 25.529 & 26.031 & 24.900 & 24.793 \\
    Optimal pair & 24.463 & 24.524 & 23.903 & 23.830 \\
    Optimal & 23.594 & 23.907 & 23.292 & 23.236 \\ \bottomrule
    \end{tabularx}
\end{subtable}

\vspace{0.25cm}
\begin{subtable}{\textwidth}
  \centering
  \caption{Weighted tanking $\mathit{WT}$}
  \label{Table3b}
  \rowcolors{1}{gray!20}{}
    \begin{tabularx}{0.9\textwidth}{lCCcC} \toprule \hiderowcolors
    \multirow{2}{*}{Group labelling rule} & \multicolumn{4}{c}{Group match scheduling policy} \\
    		 & Original & Random & Optimal uniform & Optimal \\ \bottomrule \showrowcolors
    Original & 25.456 & 28.335 & 26.184 & 26.022 \\
    Random & 27.428 & 28.360 & 26.406 & 26.221 \\
    Optimal pair & 23.456 & 24.053 & 21.966 & 21.817 \\
    Optimal & 20.968 & 22.296 & 19.990 & 19.788 \\ \bottomrule
    \end{tabularx}
\end{subtable}
\end{table}

Table~\ref{Table3} compares the two tanking measures under the three group labelling rules and the three group match scheduling policies if the group composition remains the same as given in Table~\ref{Table1}. It also shows the particular case of the 2022 FIFA World Cup, which has been a single realisation of random group labelling and scheduling. Naturally, the threat of tanking decreases if a more radical intervention is considered. However, the contribution of both optimal group labelling and scheduling remains low compared to the benefit of the two middle policies (optimal pair group labelling and optimal uniform scheduling).
To conclude, neither reform is able to substantially mitigate the risk of tanking; for instance, the probability of tanking can be reduced by less than 3 percentage points.

\begin{table}[t!]
  \centering
  \caption{Reduction of tanking by different policies in the 2022 FIFA World Cup}
  \label{Table4}
  
\begin{subtable}{\textwidth}
  \centering
  \caption{Probability of tanking $T$}
  \label{Table4a}
  \rowcolors{1}{gray!20}{}
    \begin{tabularx}{0.8\textwidth}{lCcC} \toprule \hiderowcolors
    \multirow{2}{*}{Group labelling rule} & \multicolumn{3}{c}{Group match scheduling policy} \\
    		& Random & Optimal uniform & Optimal \\ \bottomrule \showrowcolors 
    Random & 0.00\% & \textcolor{gray!20}{0}4.34\% & \textcolor{gray!20}{0}4.76\% \\
    Optimal pair & 5.79\% & \textcolor{white}{0}8.18\% & \textcolor{white}{0}8.46\% \\
    Optimal & 8.16\% & 10.52\% & 10.74\% \\ \toprule
    \end{tabularx}
\end{subtable}

\vspace{0.25cm}
\begin{subtable}{\textwidth}
  \centering
  \caption{Weighted tanking $\mathit{WT}$}
  \label{Table4b}
  \rowcolors{1}{gray!20}{}
    \begin{tabularx}{0.8\textwidth}{lCcC} \toprule \hiderowcolors
    \multirow{2}{*}{Group labelling rule} & \multicolumn{3}{c}{Group match scheduling policy} \\
    		& Random & Optimal uniform & Optimal \\ \bottomrule \showrowcolors 
    Random & \textcolor{gray!20}{0}0.00\% & \textcolor{gray!20}{0}6.89\% & \textcolor{gray!20}{0}7.54\% \\
    Optimal pair & 15.19\% & 22.55\% & 23.07\% \\
    Optimal & 21.38\% & 29.51\% & 30.23\% \\ \toprule
    \end{tabularx}
\end{subtable}
\end{table}

Table~\ref{Table4} reveals the relative performance of our reform proposals.
Scheduling is not a powerful tool against tanking, choosing the optimal match order separately for each group improves our measures by no more than 8\%, which is much smaller than the impact found by the previous literature with respect to the risk of collusion within a group \citep{ChaterArrondelGayantLaslier2021, Guyon2020a, Stronka2020}. It does not make much sense to use a non-uniform schedule across the groups.
Group labelling turns out to be a stronger policy, even in its weaker version when only the kick-off times of the fixed group pairs are optimised. Nonetheless, the risk of tanking cannot be decreased considerably without changing any fundamental aspect of the tournament design, especially if we focus on the probability of tanking.

\section{Conclusions} \label{Sec5}

The current paper has investigated sports tournaments consisting of a group stage followed by a knockout phase. If the knockout brackets are predetermined, then misaligned incentives may arise in the groups that are finished later as illustrated by several historical examples. Since it is crucial to avoid situations where deliberately conceding some goals can be beneficial for a team, we have proposed a simulation framework to quantify the threat of tanking.

The suggested approach has been applied to the 2022 FIFA World Cup. Its design could have been improved by changing the random group labelling and group match schedules but neither intervention mitigates the risk of tanking considerably. Ensuring the attractiveness and competitiveness of the matches played in the last round of the group stage seems to require a more fundamental change in the tournament format such as the following:
\begin{itemize}
\item
Dynamic scheduling when the schedule of the group matches depends on the results in the previous rounds \citep{GuajardoKrumer2023, KrumerMegidishSela2023, Stronka2024};
\item
Opponent choice when higher-ranked teams are allowed to choose their next opponents in the knockout phase \citep{Guyon2020a, HallLiu2024};
\item
Randomised tie-breaking when the order of the teams having the same number of points is decided by a biased lottery favouring teams with a superior goal difference rather than by the usual deterministic rules \citep{Stronka2024}.
\end{itemize}
This recommendation calls for more studies on sporting contests where similar policies are already used \citep{LunanderKarlsson2023}.

Hopefully, our work will inspire future research to quantify the degree of violating incentive compatibility and other desired properties of tournament designs. Such a project can explore the inevitable trade-offs between various aims of the organiser and, ultimately, improve the fairness of sports tournaments and make the matches more exciting, which will certainly increase the value of media rights.

\section*{Acknowledgements}
\addcontentsline{toc}{section}{Acknowledgements}
\noindent
This paper could not have been written without \emph{my father} (also called \emph{L\'aszl\'o Csat\'o}), who has primarily coded the simulations in Python. \\
\emph{Aleksei Y.~Kondratev} has provided useful comments and suggestions. \\
One anonymous reviewer and two colleagues have given valuable remarks on earlier drafts. \\
We are indebted to the \href{https://en.wikipedia.org/wiki/Wikipedia_community}{Wikipedia community} for summarising important details of the sports competition discussed in the paper. \\
The research was supported by the National Research, Development and Innovation Office under Grant FK 145838, and the J\'anos Bolyai Research Scholarship of the Hungarian Academy of Sciences.

\bibliographystyle{apalike}
\bibliography{All_references}

\end{document}